\documentclass[journal,twocolumn]{IEEEtran}
\usepackage{amssymb}
\usepackage{amsfonts}
\usepackage{amsmath}
\usepackage{framed}
\usepackage{algorithmic}
\usepackage[ruled,vlined]{algorithm2e}
\usepackage{graphicx,cite,multicol,multirow,diagbox,booktabs,array}
\usepackage{verbatim}
\usepackage{epstopdf}
\usepackage{subfig}
\usepackage{bm}

\usepackage[dvips]{color}
\usepackage{float}
\newtheorem{theorem}{Theorem}
\newtheorem{remark}{Remark}

\newcommand*{\QEDB}{\hfill\ensuremath{\square}}

\ifCLASSOPTIONcompsoc
\fi
\ifCLASSINFOpdf
\else
\fi
\usepackage{etoolbox}
\pretocmd{\maketitle}{%
  \markboth{}{}%
}{}{}

\begin{document}
\begin{sloppypar}

\title{Optimal Measurement of Drone Swarm in RSS-based Passive Localization with Region Constraints}

\author{Xin Cheng,~Feng Shu,~Yifan Li,~Zhihong Zhuang,~Di Wu,~and~Jiangzhou Wang,~\IEEEmembership{Fellow,~IEEE}
~\\
{\footnotesize{Xin Cheng, Yifan Li, Zhihong Zhuang are with School of Electronic and
Optical Engineering, Nanjing University of Science and Technology, Nanjing,
210094, China.}}\\

{\footnotesize{ Feng Shu is with the School of Information and Communication Engineering,
Hainan University, Haikou, 570228, China and with the School
of Electronic and Optical Engineering, Nanjing University of Science and
Technology, Nanjing, 210094, China.}} \\

{\footnotesize{Di Wu is with the School of Information and Communication Engineering,
Hainan University, Haikou, 570228, China.}}\\

{\footnotesize{Jiangzhou Wang is with the School of Engineering, University of Kent, Canterbury CT2 7NT, U.K.}}
~\\
\small{CORRESPONDING AUTHOR: DI WU and FENG SHU} 
~\\
{\footnotesize{The definitive version was published in IEEE Open Journal of Vehicular Technology, vol. 4, pp. 1-11, 2023, doi: 10.1109/OJVT.2022.3213866.}}
}

\twocolumn[
\begin{@twocolumnfalse}
\maketitle
\textbf{Abstract} Passive geolocation by multiple unmanned aerial vehicles (UAVs) covers a wide range of military and civilian applications including  rescue, wild life tracking and electronic warfare. The sensor-target geometry  is known to significantly affect the localization precision. The existing optimal sensor placement strategies  mainly work on the cases without any constraints on the sensor locations. However, UAVs cannot  fly/hover simply in arbitrary region due to realistic constraints, such as the geographical limitations, the security issues, and the max  flying speed.
In this paper, optimal  geometrical configurations of UAVs in received signal strength (RSS)-based localization  under region constraints are investigated. Employing the D-optimal criteria,  i.e., maximizing the determinant of  Fisher information matrix (FIM), such optimal problem is formulated. Based on the rigorous algebra and geometrical derivations, optimal and also closed-form configurations of UAVs under different flying states are proposed. Finally, the effectiveness and practicality of the proposed configurations are demonstrated by simulation examples.
\end{@twocolumnfalse}
]

\section{Introduction}
Passive geolocation of radio emitters, is a fundamental problem with a wide range of military and civilian applications \cite{8421288,8365918}.  Recent advancements in wireless communication and robotic technologies have made it possible to use unmanned aerial vehicles (UAVs) as aerial sensors for geolocation. Compared to traditional mediums, such as cellular localization and satellite localization, rapid deployment, flexible relocation and high chances of experiencing line-of-sight (LoS) propagation path  have been perceived as promising opportunities to provide difficult services \cite{9606574}. Due to these distinctive advantages, UAV  plays an important role in mobile
user localization, rescue, wild life tracking and electronic warfare\cite{6735682,7470933,wang2017guideloc,cliff2015online}.

Typically, UAVs in the localization task are equipped with wireless
communication modules and  appropriate sensors. Given some potentially noisy measurements from the sensors, the position of the target is estimated on UAV networks or ground servers. According to the diversity of  equipped sensors, the localization approaches can be classified into several types, including time of arrival (TOA) \cite{5714759,6184254}, time difference of arrival (TDOA) \cite{259534,7776822,8207615},  direction of arrival (DOA) \cite{8290952,9684752} and received signal strength (RSS) \cite{9606574,1247811,9360612}. Normally, the time/angle-based approach requires an open signal propagation environment  where the LoS signal  is much stronger than multi-path signals \cite{1458287}. However, the RSS-based approach requires less on the signal propagation environment by averaging the received power and modeling the complex environment \cite{1458287}. Moreover, RSS-based approach is  cost-effective since it does not require tight synchronization and calibration. In this paper, we consider a passive localization on a ground target using drone swarm equipped with RSS sensors.

Many well-known methods  have been proposed in the literature to estimate the location of radio emitters \cite{4291880,5378537,7423722}. The best possible accuracy of any unbiased  estimator is determined by the Cramer-Rao lower bounds (CRLBs).
The CRLB of the UAVs-based localization i.e., the low bound of estimation error variance, contains two ingredients: the inherent configuration  and the measurement condition.  The measurement condition is related to the signal propagation environment and  measuring property of UAVs. Besides,  the inherent configuration error is determined by UAV-target (sensor-target) geometry. It turns out that the measurement positions of UAVs significantly affect  the estimation precision. Therefore, how to determine the measurement positions of UAVs becomes an important problem.

This kind of measurement configuration problem  has attracted much attention in decades.
The CRLB matrix and the Fisher information matrix (FIM) are commonly used as the evaluation standards for designing such configurations \cite{ucinski}. Specifically, three optimal criterions have been widely used to  calculate the optimal sensor-target geometries. They are E-optimality criterion (minimizing the maximum eigenvalue of CRLB matrix), D-optimality criterion (maximizing the determinant of FIM) and A-optimality criterion (minimizing the trace of CRLB matrix). A good comparison of these criterions was presented in \cite{ucinski,9701664}. FIM  is  well-known to represent the amount of information contained in  noisy measurements, and can be expressed directly with the parameters of estimation system. Meanwhile, obtaining the CRLB matrix requires an inverse operation of  the FIM, resulting in a more complex form.  Therefore, D-optimality criterion is more suitable than the other mentioned criterions to derive analytic results, which is employed in this paper.


There is a rapidly growing research concerned with the optimal RSS sensor-target geometry problem \cite{5416784,8698828,zhao2013optimal,9701664} based on the above criterions. In \cite{5416784}, authors proposed closed-form optimal sensor-target geometry of two or three sensors with inconsistent sensor-target ranges. In \cite{8698828},  optimal geometries were acquired using a resistor network method in a three-dimension (3D) scenario.
In \cite{zhao2013optimal}, necessary and sufficient conditions of optimal placements in two-dimension (2D) and 3D scenarios were proved using frame theory. It showed that  in the equal weight case, optimal geometry of the sensors is just at the verticies of a $m$ sided regular polygon, $m$ being the number of sensors.  In
\cite{9701664}, an alternating direction method of multipliers (ADMM) framework was proposed to find the optimal sensors placement. However, these works were all limited to the optimal geometry without deployment region constraints. Note that  for UAVs in passive geolocation, region constraints can not be negligible due to such constraints as terrain, security, and max flying speed.

As to the constrained optimal sensors configuration, related studies have been published recently \cite{sun2014optimum,7551134,9057741,9684356 }.   In \cite{sun2014optimum}, the optimal sensor placement was  investigated  with some sensors being mobile and some other being stationary.   In \cite{7551134}, optimal placement problems of heterogeneous range/bearing/RSS sensors were solved under region constraints.  Three deployment regions including  a segmental arch,  a straight line, and  an external closed region were considered dependently. However, no closed-form expressions were presented for a large number of sensors.  Closed-form optimal placements of range-based sensors in a connected arbitrarily shaped region were derived in \cite{9057741}.  In \cite{9684356}, circular deployment region constraints and minimum safety distance were considered, closed-form optimal  placements were derived for a limited number of AoA sensors.
In these works \cite{sun2014optimum,7551134,9057741,9684356 }, the sensors were considered to be static once deployed. However, for UAV-enabled passive localization, considering sensors with moving ability is important.

Regarding  the dynamic sensors configuration with multiple measurements under the region constraints,  some  studies have been published \cite{6178054,7543463,8960453,9220776}.
In \cite{6178054,7543463}, UAVs    were employed to geolocate a ground target while flying towards it  from a far away area, and the trajectories were optimized at successive waypoints. In \cite{6178054},  many practical constrains  were considered, including threat/obstacle avoidance, maximum inter-UAV distance bounds and UAV turn rate constraint. A steering algorithm was proposed to update UAV trajectories. In \cite{7543463}, communication constraints were taken into account.  A leader-follower control law was designed to guide the successive movements of UAVs. In \cite{8960453,9220776}, UAVs were employed to fly over the entire AoI with potential multiple IoT devices, and the overall UAV trajectories were optimized. In \cite{8960453}, the authors proposed a novel framework based on reinforcement learning (RL) to enable a UAV  to autonomously find its trajectory that improves the localization accuracy of multiple objects in shortest time and path length, fewer waypoints, and/or lower UAV energy consumption. In \cite{9220776}, a joint position and power optimization (JPPO) framework  was proposed to optimize the UAV trajectories, and  the no-fly-zone (NFZ) and the total energy constraint were considered. However, no optimal result for overall dynamic sensors configuration has been proposed in existing works.

In this paper,  we address the problem of finding optimal geometry configuration of UAVs with moving abilities to localize a ground target using RSS measurements. Due to the  geographical limitation and the security issue, minimum flying height and minimum  UAV-target horizontal distance are considered. Each UAV is allowed to take multiple measurements below max flying speed. The max speed of UAV is also treated as a constraint. Note that the combination of the objective and the involved constraints in this paper is foundational and practical, and has not been  considered yet. Moreover, optimal solutions based on the D-optimality criterion for overall dynamic sensors configuration is proposed in this paper. Our main contributions are summarized as follows:

\begin{enumerate}
\item The optimal geometrical configuration  for drone swarm with multiple measurements in RSS-based geolocation is studied.  A special form of the FIM is derived. Combining with practical region constraints expressed mathematically, a FIM-based problem formulation is presented.
\item To solve this formulated problem,  the  optimal UAV-target geometries with fixed flying height and horizontal distance are introduced. After rigorous algebra and geometrical derivations,  optimal  solutions are provided. Altogether, four kinds of optimal configurations  are proposed corresponding to the specific flying state of UAVs, including hovering, below half circle flying, beyond half circle flying and full circle flying.
\item Extensive simulations are conducted for various testing scenarios. The simulation results demonstrate the effectiveness of proposed configurations.  Moreover, a practical scenario is simulated to verify the the practicality of these configurations as well as comparing their robustness.
\end{enumerate}

The rest of the paper is organized as follows. The system model and problem formulation are presented in Section II. In Section III, the optimal  configuration with region constraints problem is solved for hovering UAVs.  In Section IV, the solutions of such problem are derived for flying UAVs with different flying abilities, respectively.  Numerical results are presented in Section V. Finally, concluding remarks are given in Section VI.

\emph{Notations}: Boldface lower case and upper case letters denote
vectors and matrices, respectively. Sign $(\cdot)^{T}$ denotes the transpose operation and sign $\| \cdot \|$ denotes
the Frobenius norm. Sign $\mathrm{Tr}(\cdot)$ represents the trace of a matrix. Sign $\mathcal{N}$ denotes the Gaussian distribution.

\section{System Model and Problem Formulation}

\subsection{Measurement Model}
\begin{figure}
  \centering
  \includegraphics[width=0.5\textwidth]{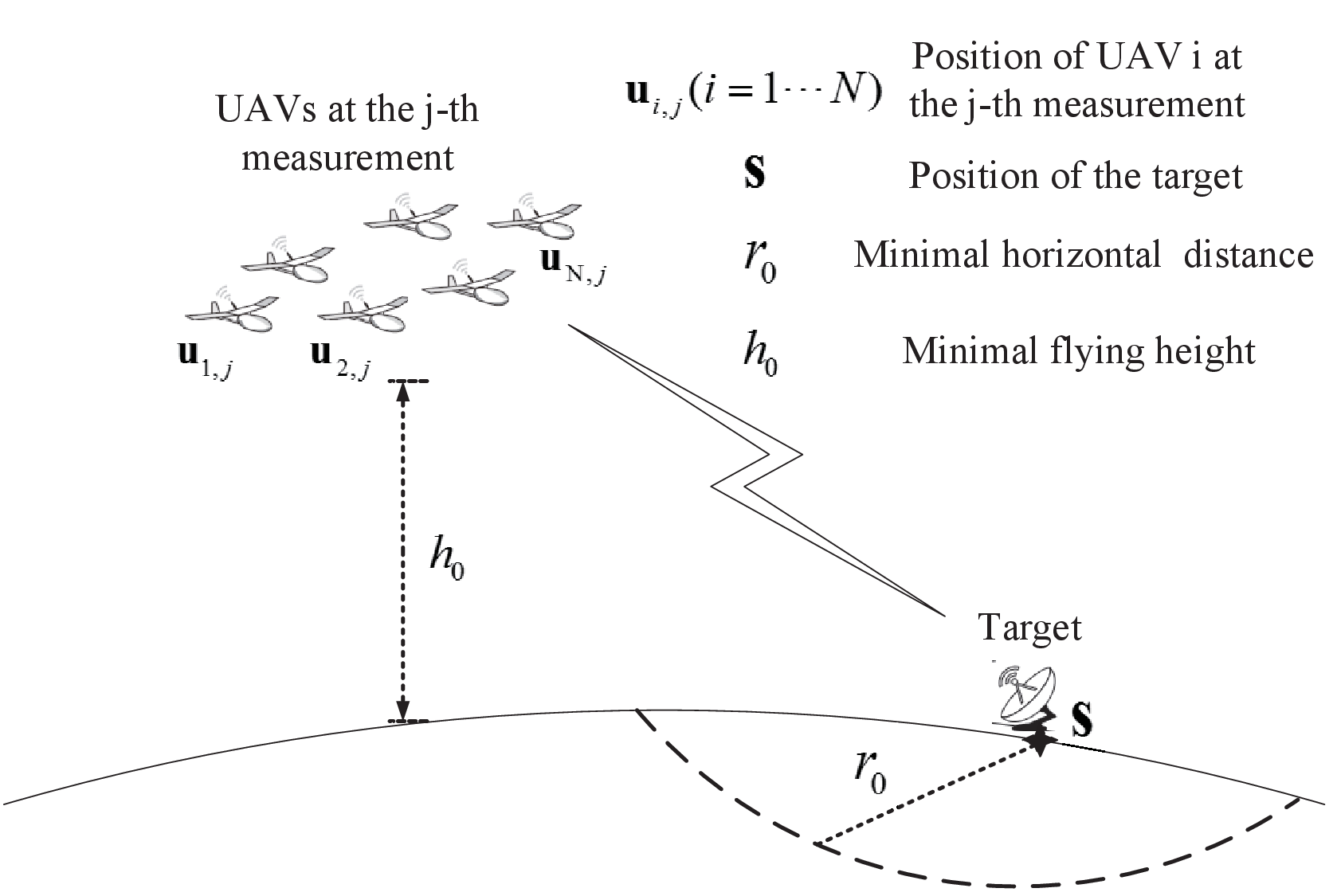}\\
  \caption{Passive localization via UAVs with region constraints.}\label{systemmodel}
\end{figure}

Consider a cooperate localization on a stationary target via UAVs, as illustrated in Fig.~\ref{systemmodel}. The target is located at an unknown position on the ground, denoted as  $\mathbf{s}=(x,y,0)$. $N$ UAVs  search  the radio-frequency (RF) signal emitted from the target and measure the RSS.

Each UAV  takes multiple measurements while flying or hovering. The measurement amount of the $i$-th UAV is denoted as $M_{i}$. The position of the $i$-th UAV at the  $j$-th  measurement  is denoted as $\mathbf{u}_{i,j}=(x_{i,j},y_{i,j},z_{i,j}),~1\leq i\leq N, 1\leq j\leq M_{i}$. Note that each UAV is aware of its own geographical position using the global positioning system (GPS). Therefore, the distance from this position  to the target can be expressed by
\begin{align}
d_{i,j}=\sqrt{(x-x_{i,j})^2+(y-y_{i,j})^2+z_{i,j}^2}, \nonumber\\
~~1\leq i\leq N,~~1\leq j\leq M_{i}.
\end{align}

According to the well-known radio propagation path loss model (in decibels) \cite{DBLP:books/cu/G2005}, the $j$-th measured RSS (in dB) of UAV $i$ can be expressed by
\begin{align}\label{RSSmodel}
R_{i,j}=\underbrace{p_{0}-10\gamma\lg(d_{i,j})}_{f_{i,j}(\mathbf{s})}+\eta_{i,j},
\end{align}
where $p_{0}$ denotes the signal power of the source and  $\gamma$ denotes the path loss exponent (PLE). It is assumed that the parameters $p_{0}$ and $\gamma$ are known (determined using calibration  or prior knowledge) \cite{6847233,6787152,9606574}. The measurement noise is denoted as $\eta_{i,j}$, and   $\eta_{i,j} \in \mathcal{N}(0,\sigma^{2}_{i})$.

Based on the measurements, $\mathbf{s}$ can be inferred using maximum likelihood (ML) estimation. Stacking all measurements of UAV  $i$ can form a $M_{i}$-dimensional column vector, shown as follows.
\begin{align}\label{RSS}
\mathbf{R}_{i}=\mathbf{f}_{i}(\mathbf{s})+\bm{\eta}_{i}.
\end{align}
Accordingly,  the probability distribution function (PDF) of the target position is given by
\begin{align}\label{localML}
&Q_{i}(\mathbf{s}) \nonumber\\
&=\frac{1}{(2\pi)^{\frac{M_{i}}{2}}|\mathbf{N}_{i}|}\exp\{-\frac{1}{2}(\mathbf{R}_{i}-\mathbf{f}_{i}(\mathbf{s}))^{T}\mathbf{N}_{i}^{-1}(\mathbf{R}_{i}-\mathbf{f}_{i}(\mathbf{s}))\},
\end{align}
where $\mathbf{N}_{i}=\sigma_{i}^{2}\mathbf{I}_{M_{i}\times M_{i}}$ is the covariance matrix of $\bm{\eta}_{i}$. As usual, measurement noises among different UAVs are independent. Therefore, the joint PDF based on the whole measurements is given by
\begin{align}\label{BigML}
Q(\mathbf{s})=\prod_{i=1}^{N}  Q_{i}(\mathbf{s}).
\end{align}

\subsection{Problem Formulation}
In this paper, we aim to find  optimal geometrical configurations in such localization task based on the D-optimality criterion. Based on the the joint PDF, the Fisher information matrix (FIM)  can be derived by
\begin{align}\label{FIM0}
\mathbf{F}=\mathbb{E}\{\nabla_{\mathbf{s}}\log{Q(\mathbf{s})}\nabla_{\mathbf{s}}\log{Q(\mathbf{s})}^{T}\}.
\end{align}

As shown in Fig.~\ref{systemmodel}, a practical scenario is considered  where  UAVs are only able to hover or fly on the limited area while measuring. Due to the safety factor (not detected by the enemy) or geographical limitations,  the minimum flying height and the minimum horizontal distance to the target must be guaranteed. Corresponding mathematical expressions are given by
\begin{align}\label{consr}
r_{i,j}=\sqrt{(x-x_{i,j})^2+(y-y_{i,j})^2}\geq  r_{0},
\end{align}
\begin{align}\label{consh}
z_{i,j}=h_{i,j}\geq  h_{0}.
\end{align}

It is assumed that all UAVs have a max flying speed, denoted as $c_{\mathrm{max}}$.   This  results in a max distance constraint between the UAV positions at two adjacent effective measurements.  Let $t_{0}$ denote the time interval of two effective measurements.  Mathematically, for the $i$-th UAV, the following condition must be satisfied.
\begin{align}\label{consspeed}
\|\mathbf{u}_{i,j}-\mathbf{u}_{i,j-1}\|\leq t_{0}c_{\mathrm{max}}.
\end{align}

Let $\mathbf{r}$ and $\mathbf{h}$ denote the collection of horizontal distances to the target and flying heights, respectively. And $\bm{\beta}$ denotes the collection of horizontal UAV-target angles where $\tan(\beta_{i,j})=\frac{x_{i,j}-x}{y_{i,j}-y}$.  It can be verified that the $\mathbf{u}_{i,j}$ can be derived from specific $r_{i,j}$, $h_{i,j}$ and $\beta_{i,j}$. To sum up, the mathematical form of this problem is as follows.
\begin{align}\label{pp1}
\mathrm{P}:&\max_{\mathbf{r},\mathbf{h},\bm{\beta}}~~~~~~~~~~~~~~~ |\mathbf{F}|    \nonumber\\
&\text{s. t.}~~~~~~~~~~~~~~(\ref{consr}), (\ref{consh}), (\ref{consspeed}).
\end{align}

\begin{remark}
Note that the objective function in the optimal measurement problem is a function with respect to the real position of the target. Unfortunately, it is unknown, otherwise, the localization task is meaningless.  However, in practice, a prior estimation may be available. Therefore, finding the optimal measurement   with respect to the prior estimation is useful to refine the estimation. We will discuss this further in the simulation part.
\end{remark}

\section{Optimal Geometrical Configurations of Hovering UAVs}
In this section, the optimal solution of  problem $\mathrm{P}$ for hovering UAVs is discussed. The hovering UAVs can be treated as flying UAVs with $c_{\mathrm{max}}=0$. In this setting, $\mathbf{u}_{i,j}=\mathbf{u}_{i}=[x_{i}~~y_{i}],~~j,=1,\cdots M_{i}$.  Firstly,  optimal UAV-target  geometries  with fixed height and horizontal distance to the target are analysed. Then, the optimal configuration with height and horizontal distance constrains is proposed.

\subsection{Optimal Geometries with Fixed Height and Horizontal Distance}
In this setting, the FIM in (\ref{bigF}), derived in  Appendix A, can be simplified to
\begin{align}\label{FIM3}
\mathbf{F}=\left(\frac{10\gamma}{\ln{10}}\right)^2\underbrace{\sum_{i=1}^{N}M_{i}\sigma_{i}^{-2}\frac{r_{i}^2}{d_{i}^4} \mathbf{g}_{i} \mathbf{g}_{i}^{T}}_{\mathbf{G}},
\end{align}
where $\mathbf{g}_{i}=\begin{bmatrix} \cos{\beta_{i}} & \sin{\beta_{i}}\end{bmatrix}$.

For the symmetric matrix $\mathbf{G}\in \mathbf{R}^{2\times2}$, the following equality is satisfied.
\begin{align}\label{fdetnorm}
\left|\mathbf{G}\right|&=\frac{1}{2}\left(\mathrm{tr}(\mathbf{G})^{2}-\mathrm{tr}(\mathbf{G}^{2})\right) \nonumber\\
&=\frac{1}{2}(\sum_{i=1}^{N}M_{i}\sigma_{i}^{-2}\frac{r_{i}^2}{d_{i}^4})^{2}-\frac{1}{2}\|\mathbf{G}\|^{2}.
\end{align}
Therefore, maximizing the $\mathrm{det}(\mathbf{F})$ is equal to minimizing the $\|\mathbf{G}\|^{2}$. It shows that
\begin{align}
\|\mathbf{G}\|^{2}&=\sum_{i=1}^{N}\sum_{j=1}^{N}M_{i}M_{j}\sigma_{i}^{-2}\sigma_{j}^{-2}\frac{r_{i}^2r_{j}^2}{d_{i}^4d_{j}^4} (\mathbf{g}_{i}^{T}\mathbf{g}_{j})^2   \nonumber\\
&\triangleq \sum_{i=1}^{N}\sum_{j=1}^{N} (c_{i}c_{j}\mathbf{g}_{i}^{T}\mathbf{g}_{j})^{2}\triangleq (\bm{\varphi}_{i}^{T}\bm{\varphi}_{j})^{2},
\end{align}
where $\bm{\varphi}=c_{i}\mathbf{g}_{i}$ and $c^{2}_{i}=M_{i}\sigma_{i}^{-2}\frac{r_{i}^2}{d_{i}^4}$.  The vectors $\{\bm{\varphi}_{i}\}_{i=1}^{n}$ form a frame in $\mathbb{R}^{2}$ \cite{4286567,4350231,casazza2006physical}. According to \cite{zhao2013optimal}, $(\bm{\varphi}_{i}^{T}\bm{\varphi}_{j})^{2}$ is just the \emph{frame potential}, and finding the minimizer of  the frame potential can be categorized as regular and irregular cases determined by the \emph{irregularity} of $\{c_{i}\}_{i=1}^{n}$. For convenience, let $\max\{ M_{i}\sigma_{i}^{-2}\frac{r_{i}^2}{d_{i}^4}| i=1,\cdots N\}=M_{k}\sigma_{k}^{-2}\frac{r_{k}^2}{d_{k}^4}$.
Referring to \cite{zhao2013optimal},  it is straightforward to derive  the specific conditions of regular and irregular cases in our scenario.  Regular case means $M_{k}\sigma_{k}^{-2}\frac{r_{k}^2}{d_{k}^4}\leq \frac{1}{2}\sum_{i=1}^{N} M_{i}\sigma_{i}^{-2}\frac{r_{i}^2}{d_{i}^4}$, while irregular case means $M_{k}\sigma_{k}^{-2}\frac{r_{k}^2}{d_{k}^4}>\frac{1}{2}\sum_{i=1}^{N} M_{i}\sigma_{i}^{-2}\frac{r_{i}^2}{d_{i}^4}$. Intuitively, a case is regular when no measuring ability of UAV (defined by $M_{i}\sigma_{i}^{-2}\frac{r_{i}^2}{d_{i}^4}$ for the $i$-th UAV) is much larger than the others. Using the frame theory \cite{4286567,4350231,casazza2006physical} as introduced in \cite{zhao2013optimal}, there exists  optimal geometries ($\{\mathbf{g}_{i}\}_{i=1}^{n}$ ) minimizing $\|\mathbf{G}\|^{2}$ in both \emph{regular} and \emph{irregular} cases. Accordingly, optimal geometries are summarized  in Theorem \ref{thm1}  and Theorem  \ref{thm2}, combining with the focused settings.

\begin{theorem}[Regular optimal geometry]
When $M_{k}\sigma_{k}^{-2}\frac{r_{k}^2}{d_{k}^4}\leq \frac{1}{2}\sum_{i=1}^{N} M_{i}\sigma_{i}^{-2}\frac{r_{i}^2}{d_{i}^4}$ (regular case), we have
\begin{align}
\|\mathbf{G}\|^{2}\geq \frac{1}{2}(\sum_{i=1}^{N}M_{i}\sigma_{i}^{-2}\frac{r_{i}^2}{d_{i}^4})^{2}.
\end{align}
The equality holds if and only if
\begin{equation}\label{ccc2}
\sum_{i=1}^{N} M_{i}\sigma_{i}^{-2}\frac{r_{i}^2}{d_{i}^4} \mathbf{g}_{i}\mathbf{g}_{i}^{T}=\frac{1}{2}  \sum_{i=1}^{N} M_{i}\sigma_{i}^{-2}\frac{r_{i}^2}{d_{i}^4} \mathbf{I}.
\end{equation}
\label{thm1}
\end{theorem}

\begin{theorem}[Irregular optimal geometry]
When $M_{k}\sigma_{k}^{-2}\frac{r_{k}^2}{d_{k}^4}>\frac{1}{2}\sum_{i=1}^{N} M_{i}\sigma_{i}^{-2}\frac{r_{i}^2}{d_{i}^4}$ (irregular case), we have
\begin{align}
\|\mathbf{G}\|^{2}\geq  (M_{k}\sigma_{k}^{-2}\frac{r_{k}^2}{d_{k}^4})^2+\left(\sum_{i=1,i\neq k}^{N}M_{i}\sigma_{i}^{-2}\frac{r_{i}^2}{d_{i}^4}\right)^{2}.
\end{align}
The equality holds if and only if
\begin{align}\label{ccc3}
\mathbf{g}_{k}^{T} \mathbf{g}_{i}=0,~~ i=1,\cdots N,~~i\neq k.
\end{align}
\label{thm2}
\end{theorem}

After trivial algebraic operations, the max information amounts in both cases are obtained, given by
\begin{align}\label{detfre}
\left|\mathbf{F}^{*}_{\mathrm{regular}}\right|=\frac{1}{4}\left(\frac{10\gamma}{\ln{10}}\right)^4(\sum_{i=1}^{N}M_{i}\sigma_{i}^{-2}\frac{r_{i}^2}{d_{i}^4})^{2},
\end{align}
\begin{align}\label{detfirre}
&\left|\mathbf{F}^{*}_{\mathrm{irregular}}\right|=\left(\frac{10\gamma}{\ln{10}}\right)^4(M_{k}\sigma_{k}^{-2}\frac{r_{k}^2}{d_{k}^4})(\sum_{i=1,i\neq k}^{N}M_{i}\sigma_{i}^{-2}\frac{r_{i}^2}{d_{i}^4}).
\end{align}

\begin{figure*}
  \centering
  \includegraphics[width=1.0\textwidth]{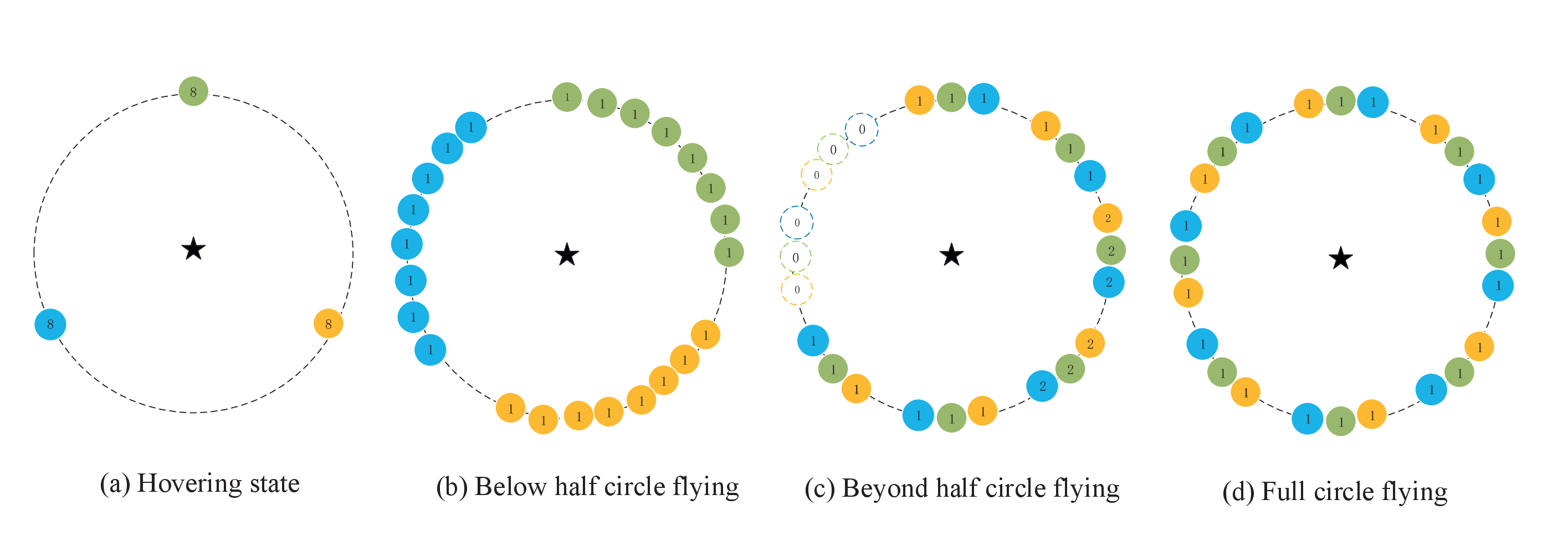}\\
  \caption{Optimal UAV-target angle configurations under various states in a regular case. ($N=3$, $M_{i}=8,~i=\cdots N$. The  star  represents the target and the solid circles represent the UAVs with different colors to distinguish them.  The numbers in the circles represent the number of measurements at the position.)}\label{opmalfigur}
\end{figure*}

\subsection{Optimal Geometrical Configurations}
Due to the split property of the max determinant of FIM, shown in formula (\ref{detfre}) and (\ref{detfirre}), it is straightforward that in both cases, the optimal horizontal distance and height of UAVs are
\begin{equation}\label{c3}
r_{i}^{\ast}=r^{\ast}=\max\{r_{0},h_{0}\},~~i=1\cdots N. \\
\end{equation}
\begin{equation}\label{c4}
h_{i}^{\ast}=h^{\ast}=h_{0},~~i=1\cdots N.
\end{equation}

In this configuration, the distance between UAVs to target are all equal to $\sqrt{(r^{\ast})^2+(h^{\ast})^2}$, denoted as $d^{\ast}$.
However, this result is only applicable to the special regular or irregular case.  To solve problem $\mathrm{P}$, further analysis is presented in the following.

For convenience, let $\max\{ M_{i}\sigma_{i}^{-2}| i=1,\cdots N\}=M_{t}\sigma_{t}^{-2}$. Besides, we define
$\varpi_{a}=M_{t}\sigma_{t}^{-2}\frac{r_{t}^2}{d_{t}^4}$,  $\varpi_{b}=\sum_{i=1,i\neq t}^{N}M_{i}\sigma_{i}^{-2}\frac{r_{i}^2}{d_{i}^4}$, $\varpi^{*}_{a}=M_{t}\sigma_{t}^{-2}\frac{(r^{\ast})^2}{(d^{\ast})^4}$, $\varpi_{b}^{*}=\sum_{i=1,i\neq t}^{N}M_{i}\sigma_{i}^{-2}\frac{(r^{\ast})^2}{(d^{\ast})^4}$
and $c=\left(\frac{10\gamma}{\ln{10}}\right)^4$. Moreover, (\ref{detfre}) is simplified as $\Psi_{1}=\frac{1}{4}c(\varpi_{a}+\varpi_{b})^2$ and (\ref{detfirre}) is simplified as $\Psi_{2}=c\varpi_{a}\varpi_{b}$.

\begin{theorem}
When $M_{t}\sigma_{t}^{-2}\leq \frac{1}{2}\sum_{i=1}^{N} M_{i}\sigma_{i}^{-2}$, the optimal configuration satisfies (\ref{c3}), (\ref{c4}) and (\ref{ccc2}).
\label{thm3}
\end{theorem}

\indent \emph{Proof:}
With (\ref{c3}), (\ref{c4}) and (\ref{ccc2}) satisfied,  $\mathrm{det}(\mathbf{F})=\Psi_{1}(\varpi^{*}_{a},\varpi^{*}_{b})$.
Obviously, $\Psi_{1}(\varpi^{*}_{a},\varpi^{*}_{b})\geq \Psi_{1}(\varpi_{a},\varpi_{b})$. Therefore, it is optimal in regular cases.
Because $\Psi_{1}(\varpi_{a},\varpi_{b})>\Psi_{2}(\varpi_{a},\varpi_{b})$ for any $\varpi_{a}$,$\varpi_{b}$,  $\Psi_{1}(\varpi^{*}_{a},\varpi^{*}_{b})\geq \Psi_{2}(\varpi_{a},\varpi_{b})$. Therefore, it is optimal compared to any irregular cases.
From above, the configuration satisfies (\ref{c3}), (\ref{c4}) and (\ref{ccc2}) is the optimal solution of problem $\mathrm{P}$.
\QEDB

\begin{theorem}
When $M_{t}\sigma_{t}^{-2}> \frac{1}{2}\sum_{i=1}^{N} M_{i}\sigma_{i}^{-2}$, the optimal configuration satisfies (\ref{c3}), (\ref{c4}) and (\ref{ccc3}).
\label{thm4}
\end{theorem}

\indent \emph{Proof:}
With (\ref{c3}), (\ref{c4}) and (\ref{ccc3}) satisfied,  $\mathrm{det}(\mathbf{F})=\Psi_{2}(\varpi^{*}_{a},\varpi^{*}_{b})$. Obviously, $\Psi_{2}(\varpi^{*}_{a},\varpi^{*}_{b})\geq \Psi_{2}(\varpi_{a},\varpi_{b})$. Therefore, the configuration is optimal in irregular cases, and
the max determinant of $\mathbf{F}$  is equal to $cM_{t}\sigma_{t}^{-2}(\sum_{i=1,i\neq t}^{N}M_{i}\sigma_{i}^{-2})\frac{(r^{\ast})^4}{(d^{\ast})^8}$.

However, there is a candidate in regular cases. Since $\Psi_{1}(\varpi_{a},\varpi_{b})$ is an increasing function with respect to $\varpi_{a}$ and  $\varpi_{b})$, $\Psi_{1}$  reaches the max value at an extreme point where $\varpi_{a}$ and $\varpi_{b}$ are max in the domain of definition.  It is easy to verify that  the max $\varpi_{a}$ and $\varpi_{b}$ satisfy $\varpi_{a}=\varpi_{b}=\sum_{i=1,i\neq t}^{N}M_{i}\sigma_{i}^{-2}\frac{(r^{\ast})^2}{(d^{\ast})^4}$.
It means that except the $t$-th UAV, the other UAVs satisfy the boundary conditions of horizontal distance and height, while the $t$-th UAV compromises with them to satisfy the regular condition.  Therefore,  the max determinant of $\mathbf{F}$ in regular cases  is equal to $c(\sum_{i=1,i\neq t}^{N}M_{i}\sigma_{i}^{-2})^2\frac{(r^{\ast})^4}{(d^{\ast})^8}$.

At last, comparing the max determinant of $\mathbf{F}$ in  two cases yields
\begin{align}
c(\sum_{i=1,i\neq t}^{N}M_{i}\sigma_{i}^{-2})^2\frac{(r^{\ast})^4}{(d^{\ast})^8}< cM_{t}\sigma_{t}^{-2}(\sum_{i=1,i\neq t}^{N}M_{i}\sigma_{i}^{-2})\frac{(r^{\ast})^4}{(d^{\ast})^8}.
\end{align}
From the above, the configuration satisfies (\ref{c3}), (\ref{c4}) and (\ref{ccc3}) is the  optimal solution of problem $\mathrm{P}$.
\QEDB

\begin{remark}
Theorem  \ref{thm3} and Theorem \ref{thm4} give  optimal configurations of hovering UAVs for all cases. The optimal distance and height configurations are straightforward. With these configurations,  UAVs are all hovering at a horizontal circle about the target.  As for the optimal UAV-target horizontal angles in the regular case,  the equality condition in Theorem~\ref{thm1} is proposed.
In order to obtain an analytic result satisfying this condition, the algorithm of constructing regular optimal placements proposed in \cite{zhao2013optimal} can be used. Combining the focused settings, it is adjusted as  Algorithm.~\ref{algorthm}. In the irregular case, the optimal angles configuration implies that UAVs from $1$ to $N$ except $t$ are collinear with the target and the line is orthogonal to the line  passing through  the $t$-th UAV and the target.
\end{remark}

\begin{figure}
  \centering
  \includegraphics[width=0.4\textwidth]{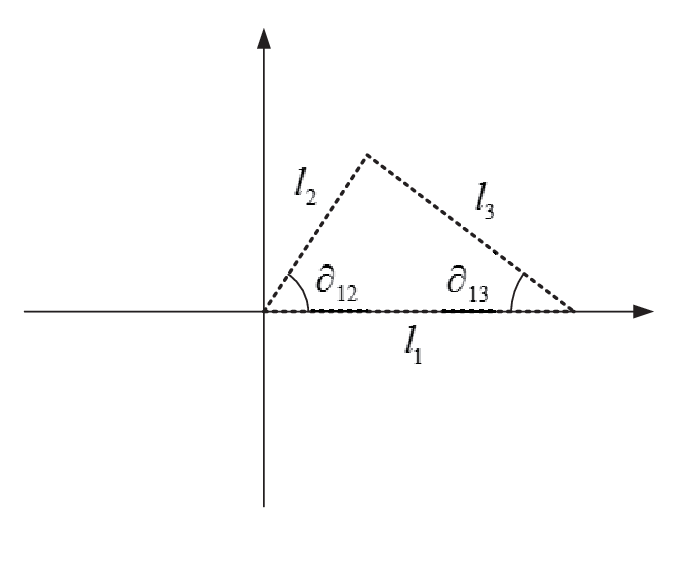}\\
  \caption{An illustration of the parameters in Algorithm.~\ref{algorthm}.}\label{figal}
\end{figure}

\begin{algorithm}
\caption{Finding optimal $\{\beta_{i}\}_{i=1}^{N}$ in the regular case.}
\label{algorthm}
\LinesNumbered
\KwIn{$\{c_{i}\}_{i=1}^{N}$ ($c^{2}_{i}=M_{i}\sigma_{i}^{-2}\frac{r_{i}^2}{d_{i}^4}$).}
\KwOut{$\{\beta_{i}\}_{i=1}^{N}$.}

Initialization: Reorder the serial number of UAVs, so that $c_{1}\geq  c_{2}\geq \cdots \geq  c_{N}$. $n_{0}=2$.

\While{$n_{0}\leq N$}{ \
                \eIf{($c_{1}^{2}+\cdots+c_{n_{0}-1}^{2}\leq\frac{1}{2} \sum_{i=1}^{N}c_{i}^{2}$)$\land$($c_{1}^{2}+\cdots+c_{n_{0}-1}^{2}+c_{n_{0}}^{2}\geq \frac{1}{2}\sum_{i=1}^{N}c_{i}^{2}$)}
                { $l_{1}=c_{1}^{2}+\cdots c_{n_{0}-1}^{2}$;$l_{2}=c_{n_{0}}^{2}$;$l_{3}=c_{n_{0}+1}^{2}+ \cdots +c_{N}^{2}$\;
                Compute interior angle $\alpha_{12}$ and $\alpha_{13}$ of the triangle with side lengths as $l_{1}$, $l_{2}$ and $l_{3}$ (See Figure. 2)\;
                Choose $\beta_{i}=0$ for $i\in\{1,\ldots, n_{0}-1\}$,  $\beta_{i}=\frac{\pi+\alpha_{12}}{2}$ for $i=n_{0}$, and $\beta_{i}=\frac{\pi-\alpha_{13}}{2}$ for $i\in\{n_{0}+1,\ldots,N\}$.
                }
                {
                  $n_{0}=n_{0}+1$.
                }　
}
\end{algorithm}

\section{Optimal Geometrical Configurations for Flying UAVs}
In this section, based on the optimal UAV-target geometry in the last section,  optimal geometrical configurations for flying UAVs  are developed.

\subsection{Below Half Circle Flying}
Assume each UAV cannot complete a half circle flying with respect to the target during the localization task, i.e.,  $t_{0}M_{i}c_{max}< \pi r^{\ast},~~i=1\cdots N$. We consider a special case where $M_{i}=M,~~i=1\cdots N$ and $\max\{\sigma_{i}^{-2}| i=1,\cdots N\}\leq\frac{1}{2}\sum_{i=1}^{N} \sigma_{i}^{-2}$ (the regular case). The FIM in (\ref{bigF}) is rewritten as
\begin{align}\label{fimeq2}
\mathbf{F}=\left(\frac{10\gamma}{\ln{10}}\right)^2\sum_{j=1}^{M}\left(\underbrace{\sum_{i=1}^{N} \sigma_{i}^{-2}\frac{r_{i,j}^2}{d_{i,j}^4} \mathbf{g}_{i,j} \mathbf{g}_{i,j}^{T}}_{\mathbf{G}_{j}}\right).
\end{align}
Note that $\mathbf{G}_{j}$ is just a special form of $\mathbf{G}$ defined in (\ref{FIM3}).  Due to the additivity of $\mathbf{F}$, it seems effective to make the UAV-target geometry at each measurement maintain the optimality. The following discussions are based on this idea.

According to Theorem \ref{thm3}, the optimal configuration in terms of  $\mathbf{G}_{j}$  satisfies (\ref{c3}), (\ref{c4}) and (\ref{ccc2}) with  $M_{i}=1$.
To satisfy (\ref{ccc2}) with  $M_{i}=1$,  Algorithm.~\ref{algorthm} can be used. Let $[\beta_{1,0}~~\beta_{2,0}~~\cdots~~\beta_{N,0}]$  denote the analytic angle configuration by Algorithm.~\ref{algorthm}. According to the principle of equivalent placements \cite{zhao2013optimal}, rotating  the overall angles around the target maintains the optimal property. Therefore  $[\beta_{1,0}+\beta~~\beta_{2,0}+\beta~~\cdots~~\beta_{N,0}+\beta]$ is  also the angle configuration satisfying (\ref{ccc2}), where $\beta$ is an arbitrary angle. Combining the flying state of UAVs, the following UAV-target horizontal angle configuration is  proposed.
\begin{align}\label{af4}
\beta_{i,j}=\beta_{i,0}+(j-1)\frac{ct_{0}}{r^{\ast}},~~0\leq c< c_{max}.
\end{align}

From the above, the following theorem is proposed.
\begin{theorem}
When $t_{0}M_{i}c_{max}< \pi r^{\ast},~M_{i}=M,~\max\{\sigma_{i}^{-2}| i=1,\cdots N\}\leq\frac{1}{2}\sum_{i=1}^{N}\sigma_{i}^{-2},~~i=1\cdots N$, an optimal configuration is (\ref{c3}), (\ref{c4}) and  (\ref{af4}).
\end{theorem}\label{thm5}

\indent \emph{Proof:}
Under the configuration satisfying (\ref{c3}), (\ref{c4}) and  (\ref{af4}), the objective function value of  problem $\mathrm{P}$ is $ \frac{1}{4}\left(\frac{10\gamma}{\ln{10}}\right)^4\left(\sum_{i=1}^{N}M\sigma_{i}^{-2}\frac{(r^{\ast})^2}{(d^{\ast})^4}\right)^{2}$. Note that this value is equal to  $|\mathbf{F}|_{3}$, i.e., the upper bound of the objective function value in problem $\mathrm{P}$,  as seen in Appendix B. Therefore, the proposed configuration is an optimal solution of problem $\mathrm{P}$.
\QEDB

\begin{remark}
When $\max\{\sigma_{i}^{-2}| i=1,\cdots N\}>\frac{1}{2}\sum_{i=1}^{N}\sigma_{i}^{-2}$ (an irregular case),  the optimal  UAV-target horizontal angle configuration at one measurement can be obtained using  Theorem \ref{thm2}. Making the UAV-target geometry  at each measurement maintain the optimality is also possible. However, in this scenario, the proposed configuration may not be optimal, thereby deserving further researches.
\end{remark}

\begin{remark}
When $0<t_{0}M_{i}c_{max}< \pi r^{\ast},~~i=1\cdots N$, the  optimal configuration in hovering state is also a selectable  optimal solution. Fig.~\ref{opmalfigur} illustrates both of them. But flying with even a slow speed is more robust  considering the prior estimation error in  practice.   Further discussions are shown in the simulation part.
\end{remark}

\subsection{Beyond Half Circle or Full Circle Flying}
Assume each UAV can complete a full horizontal circle flying with respect to the target during the localization task, i.e.,  $t_{0}M_{i}c_{\mathrm{max}}\geq 2\pi r^{\ast},~i=1\cdots N$. Let $\beta_{0,i}$ denote  the $i$-th UAV-target horizontal angle at the starting position. The FIM in (\ref{bigF}) can be rewritten as
\begin{align}
\mathbf{F}=\left(\frac{10\gamma}{\ln{10}}\right)^2\sum_{i=1}^{N}\left(\underbrace{\sum_{j=1}^{M_{i}}\sigma_{i}^{-2}\frac{r_{i,j}^2}{d_{i,j}^4} \mathbf{g}_{i,j} \mathbf{g}_{i,j}^{T}}_{\mathbf{G}_{i}}\right)
\end{align}
Note that $\mathbf{G}_{i}$ is just a special form of $\mathbf{G}$ defined in (\ref{FIM3}).  Due to the additivity of $\mathbf{F}$, it seems effective to make the UAV-target geometry for multiple measurements of each UAV  maintain the optimality. The following discussions are based on this idea.

According to Theorem \ref{thm3}, the optimal configuration in terms of $\mathbf{G}_{i}$ satisfies (\ref{c3}), (\ref{c4}) for $i$-th UAV. In this way,  $\mathbf{G}_{i}=\sigma_{i}^{-2}\frac{(r^{\ast})^2}{(d^{\ast})^4}\sum_{j=1}^{M_{i}} \mathbf{g}_{i,j} \mathbf{g}_{i,j}^{T}$. This form  belongs to the equally-weighted optimal placements \cite{1416170,zhao2013optimal}.
Accordingly, when $M_{i}=2$, the right angle structure is optimal in terms of $\mathbf{G}_{i}$, given by
\begin{align}\label{af1}
\beta_{i,j}=\beta_{0,i}+\frac{\pi (j-1)}{2},~~j=1\cdots M_{i}.
\end{align}
When $M_{i}>2$, the uniform angular array (UAA) is optimal in terms of $\mathbf{G}_{i}$, given by
\begin{align}\label{af2}
\beta_{i,j}=\beta_{0,i}+\frac{2\pi (j-1)}{M_{i}},~~j=1\cdots M_{i}.
\end{align}

Moreover,  their exists other optimal geometrical configurations in terms of $\mathbf{G}_{i}$ for $M_{i}>2$. Let $K_{i}$ denote an arbitrarily integer number satisfying $K_{i}<M_{i}$, and $K_{i}\geq\frac{M_{i}}{2}$.
Via flipping some  positions in UAA  that are measured after the $K_{i}$-th measurement about the target, the configuration is completed.  Mathematically, in this configuration,
\begin{equation}\label{af3}
\beta_{i,j}=\left\{
	\begin{aligned}
	\beta_{0,i}+\frac{2\pi (j-1)}{M_{i}} \quad j\leq K_{i},\\
    \beta_{0,i}+\frac{2\pi (j-1)}{M_{i}}-\pi \quad K_{i}<j\leq M_{i}.\\
	\end{aligned}
	\right
	.
\end{equation}
According to the principle of equivalent placements \cite{zhao2013optimal},  this configuration is an equivalent form to the optimal full circle flying configuration, i.e, UAA.

Based on the above, the following theorem is proposed.
\begin{theorem}
When $t_{0}M_{i}c_{\mathrm{max}} \geq 2\pi r^{\ast},~M_{i}>2,~~i=1\cdots N$, an optimal configuration is (\ref{c3}), (\ref{c4}) and (\ref{af2}). Besides,   configurations satisfying (\ref{c3}), (\ref{c4}) and  (\ref{af3}) are also optimal.
\end{theorem}

\indent \emph{Proof:}
Under the configuration satisfying (\ref{c3}), (\ref{c4}) and (\ref{af2})/(\ref{af3}), the objective function value of  problem $\mathrm{P}$ is $ \frac{1}{4}\left(\frac{10\gamma}{\ln{10}}\right)^4\left(\sum_{i=1}^{N}M_{i}\sigma_{i}^{-2}\frac{(r^{\ast})^2}{(d^{\ast})^4}\right)^{2}$. Note that this value is equal to  $|\mathbf{F}|_{3}$, i.e., the upper bound of the objective function value in problem $\mathrm{P}$,  illustrated in Appendix B. Therefore, the proposed configurations are optimal solutions of problem $\mathrm{P}$.
\QEDB

\begin{remark}
By observing the objective function values under proposed optimal configurations in both hovering and flying cases, two conclusions are summarized as follows. In the regular case, in terms of the max information amount of noisy measurements,  flying state is the same as  hovering state. In the irregular case,  in terms of the max information amount of noisy measurements, flying state is over the hovering state.
\end{remark}

\begin{remark}
For $t_{0}M_{i}c_{max} \geq 2\pi r^{\ast},~M_{i}>2,~~i=1\cdots N$, the optimal configurations with below half circle flying as well as hovering state are also selectable optimal solutions. Fig.~\ref{opmalfigur} illustrates all of them. But a full circle flying is the most robust considering the prior estimation error in  practice. Further discussions are shown in the simulation part.
\end{remark}

\begin{figure}
  \centering
  \includegraphics[width=0.4\textwidth]{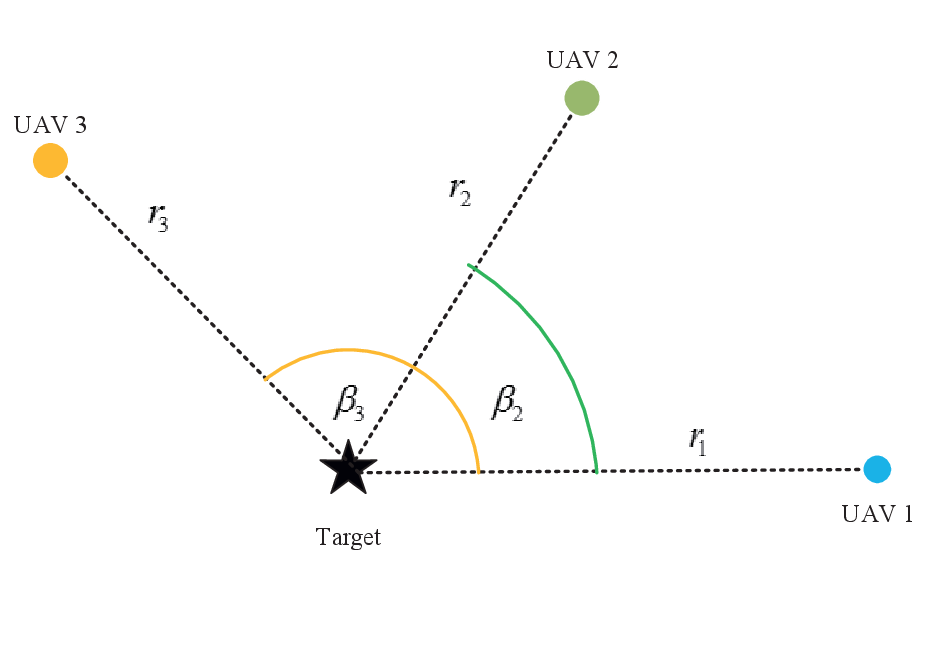}\\
  \caption{An illustration of passive localization via three hovering UAVs. (The solid circles represent
horizontal projections of UAVs.  $\beta_{i}$ represents the UAV-target horizontal angle of UAV $i$, and  $\beta_{1}=0$.)}\label{aufig}
\end{figure}

\begin{figure*}
	\centering
	\subfloat[Regular case (a).]{\label{ref}\includegraphics[width=0.45\textwidth]{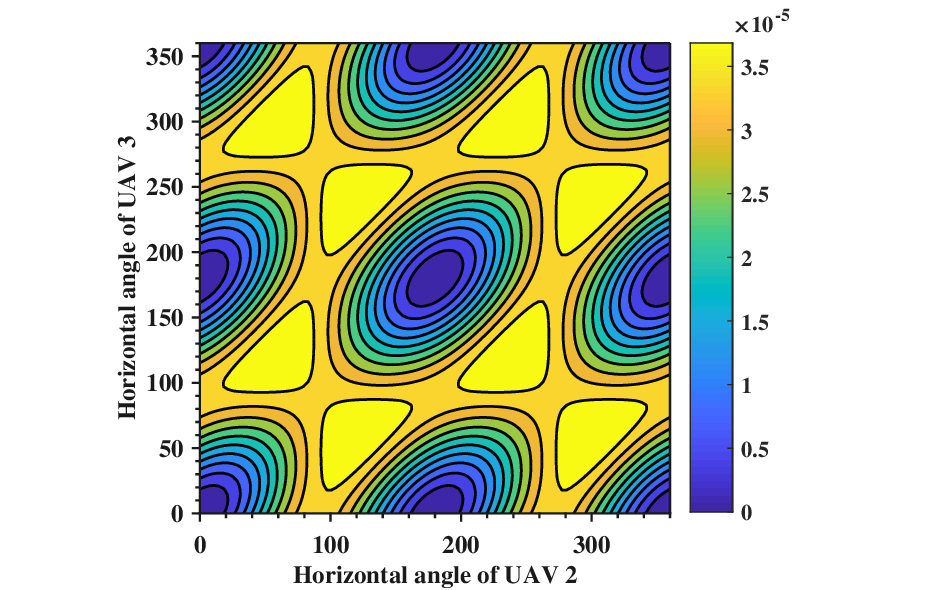}}
	\subfloat[Regular case (b).]{\label{ref}\includegraphics[width=0.45\textwidth]{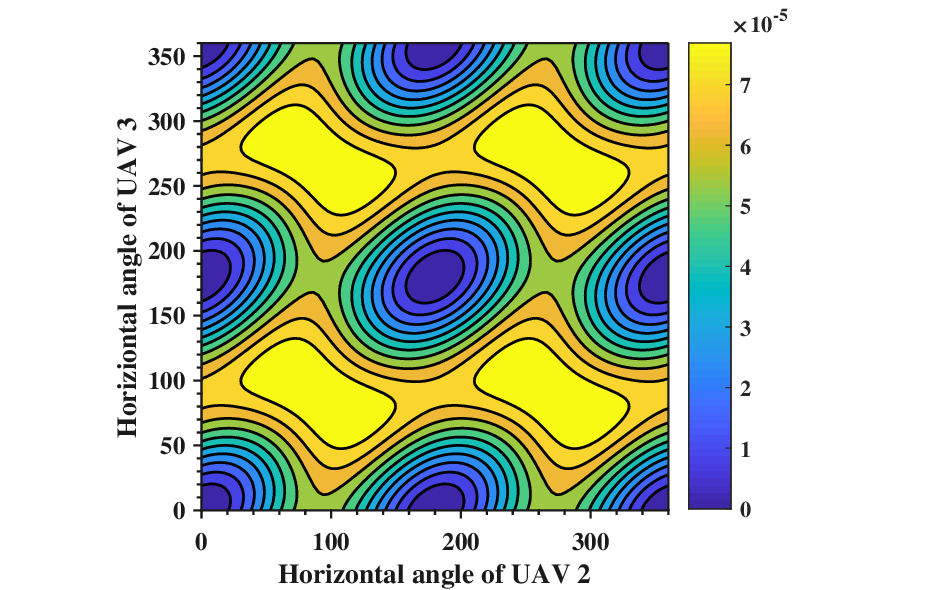}} \\
	\subfloat[Irregular case (c).]{\label{irref}\includegraphics[width=0.45\textwidth]{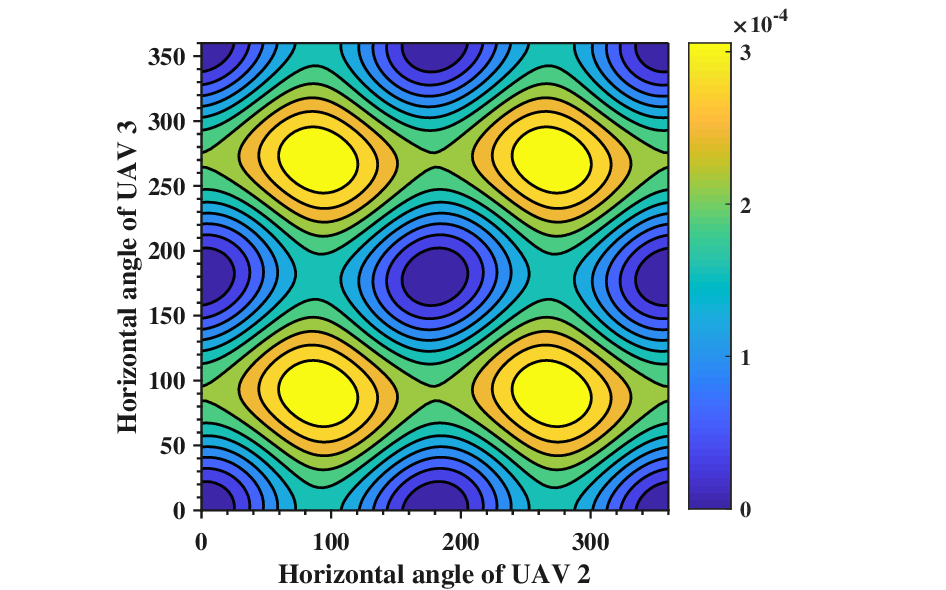}}
	\subfloat[Irregular case (d).]{\label{irref}\includegraphics[width=0.45\textwidth]{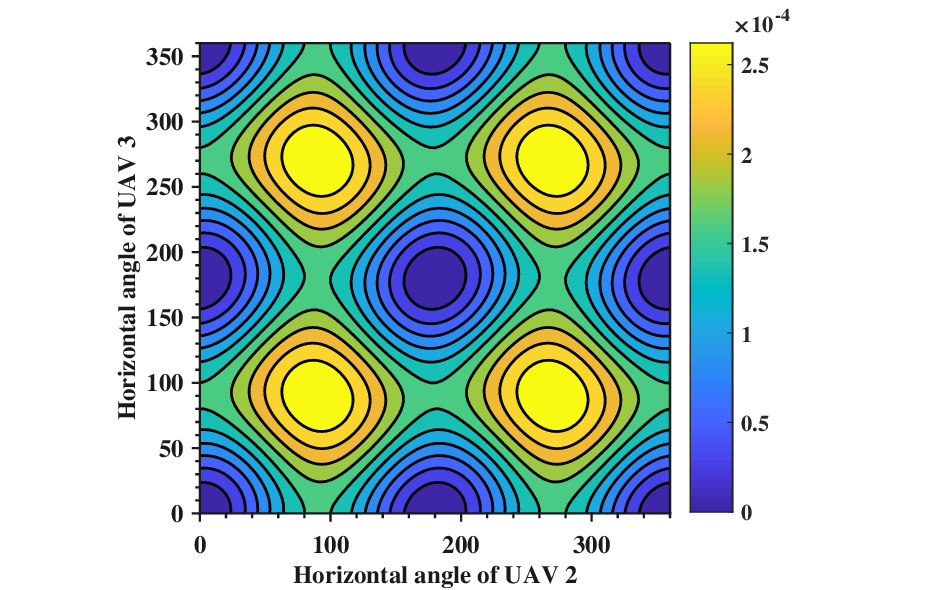}}
	\caption{Determinant of FIM via angle configurations.}
    \label{sigeometry}
\end{figure*}

\begin{figure}
	\centering
	\subfloat[$\mathrm{det}(\mathbf{F})$ via  horizontal distances  of UAVs.]{\label{fdevice}\includegraphics[width=0.5\textwidth]{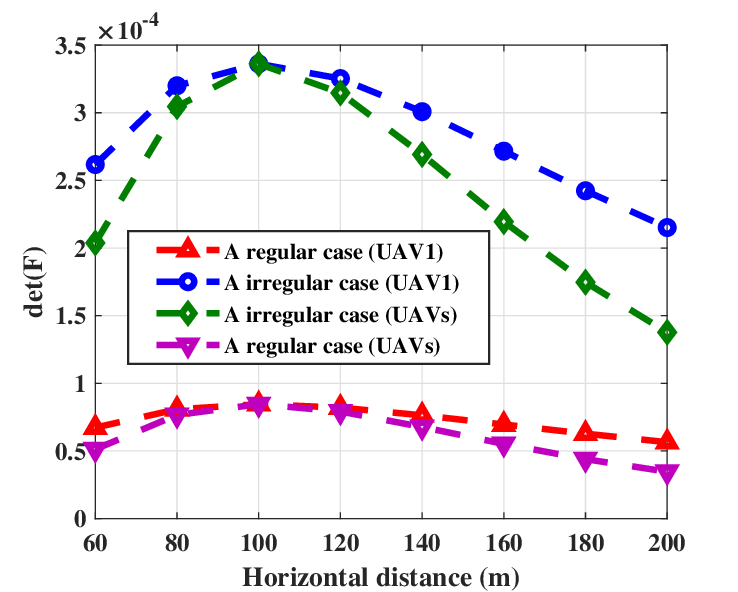}} \\
	\subfloat[$\sqrt{\mathrm{Tr}(\mathrm{CRLB})}$ via horizontal distances  of UAVs.]{\label{ftime}\includegraphics[width=0.5\textwidth]{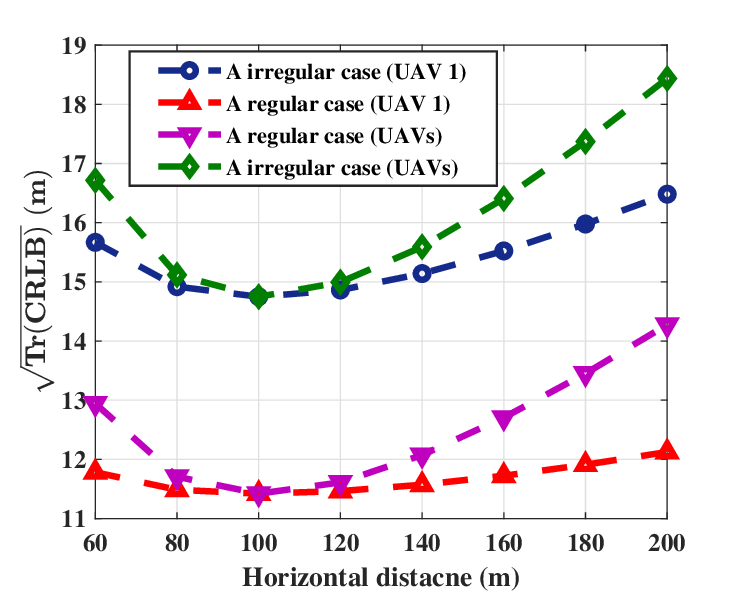}}
	\caption{Effects of the horizontal distance.}
    \label{siradiaus}
\end{figure}
\section{Simulation}
In this section, we provide numerical results to illustrate the optimal UAV-target geometry as well as the proposed optimal geometrical configurations.
Unless noted otherwise, some parameters  are set as follows: $\gamma=3$, $h_{0}=100~\mathrm{m}$, $r_{0}=60~\mathrm{m}$ and $M_{i}=16,~~i=1\cdots N$.

\subsection{Optimal UAV-Target Geometry}
In this subsection, $N=3$ UAVs equipped with RSS sensors are deployed to measure the RF signal from the target  while hovering. The horizontal distances to the target and  heights satisfy (\ref{c3}) and (\ref{c4}) respectively. The $1$-th UAV is hovering with  $\beta_{1}=0$.  Both regular and irregular cases are considered. The setting for the regular cases  are  $\sigma^{2}_{1}=16~\mathrm{dB}; \sigma^{2}_{2}=16~\mathrm{dB}; \sigma^{2}_{3}=16~\mathrm{dB}$, named as case (a) and
$\sigma^{2}_{1}=8~\mathrm{dB}; \sigma^{2}_{2}=12~\mathrm{dB}; \sigma^{2}_{3}=16~\mathrm{dB}$, named as case (b). Meanwhile, the setting for the irregular cases  are  $\sigma^{2}_{1}=2~\mathrm{dB}; \sigma^{2}_{2}=8~\mathrm{dB}; \sigma^{2}_{3}=16~\mathrm{dB}$, named as case (c) and
$\sigma^{2}_{1}=2~\mathrm{dB}; \sigma^{2}_{2}=16~\mathrm{dB}; \sigma^{2}_{3}=16~\mathrm{dB}$, named as case (d). The UAV positions are illustrated in Fig.~\ref{aufig}.

Fig.~\ref{sigeometry} shows the the FIM determinant via the UAV-target angle configuration. In the regular case (a) (a equally-weighted placement),  the optimal angle configurations are  $[60^{\circ},~120^{\circ}]$, $[60^{\circ},~300^{\circ}]$, $[120^{\circ},~60^{\circ}]$, $[120^{\circ},~240^{\circ}]$, $[240^{\circ},~120^{\circ}]$, $[240^{\circ},~300^{\circ}]$, $[300^{\circ},~60^{\circ}]$, $[300^{\circ},~240^{\circ}]$.
In the regular case (b), the optimal angle configurations are around $[103^{\circ},~72^{\circ}]$, $[103^{\circ},~252^{\circ}]$, $[283^{\circ},~102^{\circ}]$, $[283^{\circ},~253^{\circ}]$. It can be verified that all the optimal configurations are consistent with Theorem \ref{thm1}. In the irregular cases (c) and (d), the optimal angle configurations are  $[90^{\circ},~90^{\circ}]$, $[90^{\circ},~270^{\circ}]$, $[270^{\circ},~90^{\circ}]$, $[270^{\circ},~270^{\circ}]$. Obviously, the results are the same as  Theorem \ref{thm2}.

\subsection{Optimal Horizontal Distance}
In this subsection,  $N=3$ UAVs are deployed to measure the RF signal from the target while hovering. Basic settings are the same as that in  the last subsection. Both regular and irregular cases are considered. As for each case, the variance settings maintain the same as the last subsection (case (b) and case (c)). The optimal UAV-target angle configurations are used.

In Fig.~\ref{siradiaus}, the determinant of FIM  and a low bound of root mean square error (RMSE) is simulated vias  the hovering horizontal distance to the target.  Specifically, the low bound is $\sqrt{\mathrm{Tr}(\mathrm{CRLB})}$. From the figure, the following conclusions are obtained. When UAV 2 and 3 are fixed, i.e., $r_{2}=r_{3}=h_{0}$, the optimal horizontal distance of UAV 1 is  equal to $h_{0}$ ($\max\{r_{0},h_{0}\}$). When no UAV is fixed, the optimal horizontal distance of UAVs are also equal to $h_{0}$. It is concluded that the optimal UAV-target elevation angle is $45^{\circ}$ when $r_{0}\leq h_{0}$. Note that it is different to normal 2D cases  where sensors should be  close to the target. The reason is that an additional dimensional exists while measuring the target on the ground via UAVs.

\subsection{Optimal Configurations with Prior Estimation}

In this subsection, a practical scenario is considered where  the proposed  configurations are used to refine the prior estimation. Note that the proposed configurations are optimal in ideal scenario where the target position is known. $N=15$ UAVs are deployed. Among them, the measurement noise variance of ten UAVs are set to be $12~\mathrm{dB}$ and the others are set to be  $16~\mathrm{dB}$.

\begin{figure}
  \centering
  \includegraphics[width=0.5\textwidth]{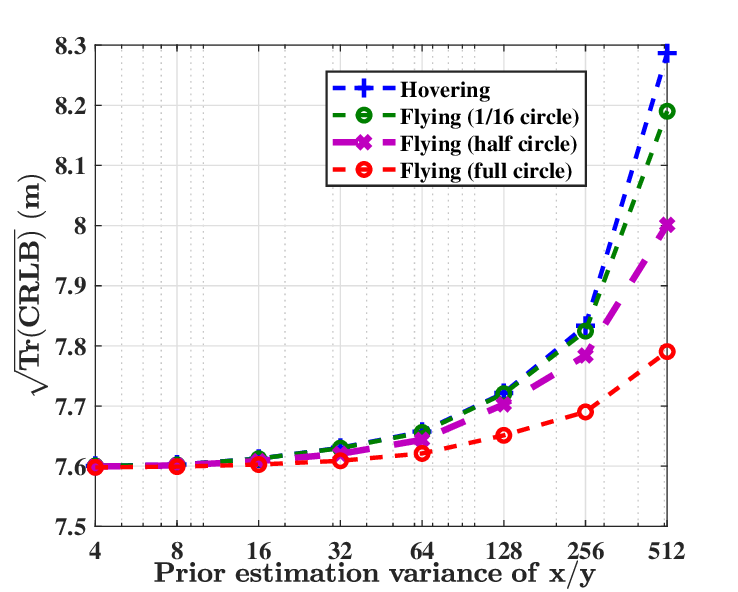}\\
  \caption{$\sqrt{\mathrm{Tr}(\mathrm{CRLB})}$ via prior estimation error variance.}\label{tuprior}
\end{figure}

Fig.~\ref{tuprior} shows the low bound of RMSE, i.e., the best estimation precision via the variance of prior estimation error under proposed configurations. Additionally, the simulation results conducted in the corresponding ideal scenario show that the best estimation precisions of proposed configurations are all equal to $7.6~\mathrm{m}$. It can be seen from the figure that even the prior estimation error variance becomes vary large, the proposed configurations still have  considerable performances. For example, when the prior estimation variance of $\mathbf{x}/\mathbf{y}$ is large as $256$, the  best estimation precisions  under proposed configurations are  less than $7.9~\mathrm{m}$, which is only $4\%$ higher than the ideal scenario. Therefore, in practical scenario, using these proposed configurations based on a prior estimation can refine the  estimation significantly. Moreover, as observed, the robustness level is ordered as follows: full circle flying, beyond half circle flying, below half circle flying, hovering.


\section{Conclusions}
In this paper, we have studied the problem of finding the optimal geometrical configuration of drone swarm to localize a ground target. Based on the optimal UAV-target geometry, optimal  configurations of UAVs have been proposed for both hovering and flying states with practical  region constraints. The numerical results have demonstrated significant benefits of the proposed configurations in both ideal and realistic examples.

\appendices
\section{A form of FIM in Multi-UAVs RSS-based Localization}
In this section, a special form of the FIM of the multi-UAVs RSS-based measurements is derived.

Referring to \cite{DBLP:journals/spl/XuOZ19}, the FIM on the RSS data measured by the $i$-th UAV ($\mathbf{F}_{i}$) can be written as
\begin{align}\label{FIMsingle}
\mathbf{F}_{i}&=\begin{bmatrix} \mathbf{a}_{x,i}^{T}\mathbf{N}_{i}^{-1}\mathbf{a}_{x,i}  &  \mathbf{a}_{x,i}^{T}\mathbf{N}_{i}^{-1}\mathbf{a}_{y,i}  \\ \mathbf{a}_{x,i}^{T}\mathbf{N}_{i}^{-1}\mathbf{a}_{y,i}  & \mathbf{a}_{y,i}^{T}\mathbf{N}_{i}^{-1}\mathbf{a}_{y,i} \end{bmatrix},
 \end{align}
where
\begin{align}
\mathbf{a}_{x,i}= \begin{bmatrix}  \frac{10\gamma}{\ln{10}} \frac{x_{i,1}-x}{d_{i,1}^2} &\frac{10\gamma}{\ln{10}} \frac{x_{i.2}-x}{d_{i,2}^2} & \cdots & \frac{10\gamma}{\ln{10}} \frac{x_{i,M_{i}}-x}{d_{i,M_{i}}^2}\end{bmatrix},
\end{align}
\begin{align}
\mathbf{a}_{y,i}=\begin{bmatrix} \frac{10\gamma}{\ln{10}} \frac{y_{i,1}-y}{d_{i,1}^2} &\frac{10\gamma}{\ln{10}} \frac{y_{i,2}-y}{d_{i,2}^2} &\cdots& \frac{10\gamma}{\ln{10}} \frac{y_{i,M_{i}}-y}{d_{i,M_{i}}^2}\end{bmatrix}.
\end{align}

Because measurement noises among different UAVs are assumed to be dependent, (\ref{FIM0}) can be expressed by
\begin{align}\label{FIM1}
\mathbf{F}
&=\mathbb{E}\{(\sum_{i=1}^{N}\nabla_{\mathbf{s}}\log{Q_{i}(\mathbf{s})})   (\sum_{i=1}^{N}\nabla_{\mathbf{s}}\log{Q_{i}(\mathbf{s})}^{T})\}
=\sum_{i=1}^{N}\mathbf{F}_{i}.
\end{align}

Substituting (\ref{FIMsingle}) into  (\ref{FIM1}) yields
\begin{align}\label{bigF}
&\mathbf{F} \nonumber\\
&=\sum_{i=1}^{N}  \begin{bmatrix} \sum_{j=1}^{M_{i}} \sigma_{i}^{-2} \mathbf{a}_{x,i}^2(j) &  \sum_{j=1}^{M_{i}} \sigma_{i}^{-2}\mathbf{a}_{x,i}(j)\mathbf{a}_{y,i}(j) \\  \sum_{j=1}^{M_{i}} \sigma_{i}^{-2} \mathbf{a}_{x,i}(j)\mathbf{a}_{y,i}(j)  &  \sum_{j=1}^{M_{i}} \sigma_{i}^{-2} \mathbf{a}_{y,i}^2(j)  \end{bmatrix} \nonumber\\
&=\left(\frac{10\gamma}{\ln{10}}\right)^2\sum_{i=1}^{N}\sum_{j=1}^{M_{i}}\sigma_{i}^{-2}\frac{r_{i,j}^2}{d_{i,j}^4} \mathbf{g}_{i,j} \mathbf{g}_{i,j}^{T},
\end{align}
where
\begin{align}
\mathbf{g}_{i,j}=\begin{bmatrix} \cos{\beta_{i,j}} & \sin{\beta_{i,j}}\end{bmatrix},
\end{align}
where $\beta_{i,j}$ is an UAV-target horizontal angle. Mathematically, $\tan(\beta_{i,j})=\frac{x_{i,j}-x}{y_{i,j}-y}$.

\section{An Upper Bound of The Objective Function Value in Problem $\mathrm{P}$}
According to  Sec.~III.A, with fixed height and horizontal distance, the max value of $|\mathbf{F}|$ in (\ref{bigF}) must fall into one of the following two values, denoted as $|\mathbf{F}|_{1}$ and $|\mathbf{F}|_{2}$,
\begin{align}\label{value1}
|\mathbf{F}|_{1}=\frac{1}{4}\left(\frac{10\gamma}{\ln{10}}\right)^4\left(\sum_{i=1}^{N}\sum_{j=1}^{M_{i}}\sigma_{i}^{-2}\frac{r_{i,j}^2}{d_{i,j}^4}\right)^{2},
\end{align}
\begin{align}\label{value2}
|\mathbf{F}|_{2}=&\left(\frac{10\gamma}{\ln{10}}\right)^4\left(\sigma_{p}^{-2}\frac{r_{p,q}^2}{d_{p,q}^4}\right)   \nonumber\\
&\left(\sum_{i=1}^{N}\sum_{j=1}^{M_{i}}\sigma_{i}^{-2}\frac{r_{i,j}^2}{d_{i,j}^4}-\sigma_{p}^{-2}\frac{r_{p,q}^2}{d_{p,q}^4}\right),
\end{align}
where $\sigma_{p}^{-2}\frac{r_{p,q}^2}{d_{p,q}^4}=\max\{\sigma_{i}^{-2} \frac{r_{i,j}^2}{d_{i,j}^4}|i,j\in\Omega\}$, and $\Omega$ is the set consisting of all values of $i,j$.

Then, consider a problem from problem $\mathrm{P}$ without the speed constrains as follows.
\begin{align}
\mathrm{P_{1}}:&\max_{\mathbf{r},\mathbf{h},\bm{\beta}}~~~~~~~~~~~~~~~ |\mathbf{F}|    \nonumber\\
&~~\text{s. t.}~~~~~~~~~~~~~(\ref{consr}), (\ref{consh}).
\end{align}
Following   Sec.~III.B, based on $|\mathbf{F}|_{1}$ and $|\mathbf{F}|_{2}$,  the max value of $|\mathbf{F}|$ in problem $\mathrm{P_{1}}$, denoted as $|\mathbf{F}^{\ast}|_{\mathrm{P_{1}}}$, must fall into one of the following two values, denoted as $|\mathbf{F}|_{3}$ and $|\mathbf{F}|_{4}$.
\begin{align}\label{value3}
|\mathbf{F}|_{3}=\frac{1}{4}\left(\frac{10\gamma}{\ln{10}}\right)^4\left(\sum_{i=1}^{N}M_{i}\sigma_{i}^{-2}\frac{(r^{\ast})^2}{(d^{\ast})^4}\right)^{2}.
\end{align}
\begin{align}\label{value4}
|\mathbf{F}|_{4}=&\left(\frac{10\gamma}{\ln{10}}\right)^4\left(\sigma_{p}^{-2}\frac{(r^{\ast})^2}{(d^{\ast})^4}\right) \nonumber\\
&\left(\left(\sum_{i=1}^{N}M_{i}\sigma_{i}^{-2}\frac{(r^{\ast})^2}{(d^{\ast})^4}\right)-\sigma_{p}^{-2}\frac{(r^{\ast})^2}{(d^{\ast})^4}\right).
\end{align}
It is easy to verify that $|\mathbf{F}|_{4} \leq  |\mathbf{F}|_{3}$. Therefore,  $|\mathbf{F}^{\ast}|_{\mathrm{P_{1}}}\leq|\mathbf{F}|_{3}$.
Let the max value of $|\mathbf{F}|$ in problem $\mathrm{P}$ denote as $|\mathbf{F}^{\ast}|_{\mathrm{P}}$.
Since  $|\mathbf{F}^{\ast}|_{\mathrm{P}}\leq |\mathbf{F}^{\ast}|_{\mathrm{P_{1}}}$, we obtain that $|\mathbf{F}^{\ast}|_{\mathrm{P}}\leq |\mathbf{F}|_{3}$. Therefore, the upper bound
 of the objective function value in problem $\mathrm{P}$ is $|\mathbf{F}|_{3}$.

\ifCLASSOPTIONcaptionsoff
\newpage
\fi

\bibliographystyle{IEEEtran}
\bibliography{IEEEfull,cite}

\end{sloppypar}
\end{document}